\documentclass[article,aps]{revtex4}
\usepackage{graphicx}% Include figure files
\usepackage{dcolumn}% Align table columns on decimal point
\usepackage{bm}% bold math

\newcommand{\be}{\begin{equation}}
\newcommand{\ee}{\end{equation}}
\newcommand{\bea}{\vspace{0.25cm}\begin{eqnarray}}
\newcommand{\eea}{\end{eqnarray}}

\def\PRL{{Phys. Rev. Lett.} }
\def\PRA{{Phys. Rev.} A }

\begin{document}

\title{ Experimental realization of a measurement conditional
unitary operation at single photon level and application to
detector characterization. }

\author{ M. Genovese}
\author{ G. Brida}
\author{ M. Gramegna}
\author{  M.L. Rastello}

 \affiliation{ Istituto
Elettrotecnico Nazionale Galileo Ferraris, Strada delle Cacce 91,
10135 Torino, Italy}

 \author{ M. Chekhova}
 \author{ L. Krivitsky}
\author{S. Kulik}

\affiliation{ Phys. Dep., M.V. Lomonosov Moscow State Univ.,
119992 Moscow, Russia. }
%\authorinfo{ genovese@ien.it. Tel. 39 011 3919253, fax 39 011 3919259, http://www.ien.it/~genovese/marco.html}

% \pagestyle{plain}
%>>>> uncomment following to start page numbering at 301
%\setcounter{page}{301}

%\maketitle

\begin{abstract}

Our last experimental results on the realization of a
measurement-conditional unitary operation at single photon level
are presented. This gate operates by rotating by $90^o$ the
polarization of a photon produced by means of Type-II Parametric
Down Conversion conditional to a polarization measurement on the
correlated photon. We then propose a new scheme for measuring the
quantum efficiency of a single photon detection apparatus by using
this set-up. We present experimental results obtained with this
scheme compared with {\it traditional} biphoton calibration. Our
results show the interesting potentiality of the suggested scheme.

\end{abstract}
 \maketitle
\section{Introduction}

The possibility of performing a unitary transformation on a qubit
conditionally on the result of the measurement on another one is a
fundamental tool for Quantum Information \cite{NC}. For example,
in teleportation protocols \cite{teleth}, teleportation is
obtained by Bob doing an opportune unitary transformation on its
sub-part of an entangled system after having received  classical
information on the result of a Bell measurement performed by Alice
on her sub-part of the entangled system and the unknown state to
be teleported. Similarly, in entanglement swapping protocols a
joint measurement of two members of two distinct entangled pairs
allows, by means of conditional unitary transformation, to create
 a desired entangled state between the two surviving
members.

 Nevertheless, in most of experiments realized up to now with
 single photons both for teleportation \cite{tele} and swapping
 \cite{swex}, this part of the
 protocol was not accomplished, and only correlation measurements with a fixed polarization selection
 were obtained. The only exception is Ref.~\cite{deMartini}, where a Pockels cell was used
 for conditional transformations in an active teleportation protocol. Conditional transformations were
 also performed in experiments where one of the photons of an entangled pair
triggered polarization rotation on the other one, to increase the
efficiency of the parity
 check~\cite{pitt1} and to produce signal photons on 'pseudodemand', i.e., in a given time interval
 after the trigger detection event~\cite{pitt2}.

In this paper we present the realization of a set-up addressed to
perform conditional  unitary transformations on the polarization
state of a photon  belonging to a polarization-entangled pair.
This experiment  is the first demonstration of 'purifying' the
polarization state of a photon that is initially prepared in a
 mixed polarization state (initial degree of polarization is equal to zero).
 As a result of transformations, the degree of polarization increases; its final value is given
 by the quantum efficiency of the 'trigger' detector. Due to this
 fact, one can notice another important advantage of our experiment: it can be used for the absolute calibration of
 photodetectors (absolute measurement of quantum efficiency), a property with large interest for various applications.
In fact, in the recent years single photon detectors have found
various important scientific and technological applications, which
demand for a precise determination of the quantum efficiency of
these apparatuses. Among them the studies about foundations of
quantum mechanics \cite{au} (and in particular Bell Inequalities
measurements \cite{bell,typeI}), quantum cryptography \cite{QCr},
quantum computation \cite{QCo,NC}, etc.

Classical calibration schemes are based on the use of a strongly
attenuated source whose (unattenuated) intensity has been
 measured by means of a radiometer. The precision of this
kind of measurement is limited by the uncertainty in the
calibration of the high insertion loss required for reaching
single photon level.

An alternative is offered by the use of photons produced by means
of parametric down conversion (PDC). These states have the
property that photons are emitted in pairs strongly correlated in
direction, wavelength and polarization. Furthermore, photon of the
same pair are emitted within tens of femtoseconds. Since the
observation of a photon on a certain direction (signal) implies
the presence of another on the conjugated direction (idler), if
this last is not observed this depends on the non-ideal quantum
efficiency of the detector, which can be measured in this way
\cite{bp1,alan}. This method is now approaching a metrological
level.

A comparison between this last scheme and our new one will be
presented in the following.

\section{ Description of the measurement-conditioned unitary gate }

Our scheme consists in measuring the polarization of a photon
belonging to a polarization entangled pair and to perform a
unitary operation, by means of a Pockels cell, on the second
member of the pair conditional to the result of this measurement.

In order to realize this set-up, first of all one needs to produce
polarization entangled states of photons by using parametric
fluorescence.

This state can be generated either by using type-II PDC
\cite{bell} or the superposition of two type I PDC emissions
\cite{typeI}, obtaining the Bell state:

\begin{equation}
\vert \psi ^+ \rangle = {\frac{ \vert H \rangle \vert V \rangle +
\vert V \rangle \vert H \rangle }{\sqrt {2}}} \label{Psi}
\end{equation}
where $H,V$ denote horizontal and vertical polarization
respectively (other phases between the two components can also be
obtained).

Then, one of the photons of the pair is addressed, after a
polarization selection, to a first detector.

When this photon (signal) is detected, we modify the polarization
of the delayed second photon (idler) of the pair by means of a
Pockels cell driven by a high voltage supply controlled by the
output of the first detector.

In particular a rotation of $90^o$ of the polarization of the
second photon can be produced after the first photon is detected
after a specific polarization selection. This realizes the desired
measurement-conditional unitary transformation.

The result of this unitary operation is that the state of the
idler photon, which was initially mixed, becomes partly purified
due to this transformation. In terms of polarization properties,
light became partly polarized (polarization degree became equal to
the quantum efficiency of the trigger detector) while initially it
was completely non-polarized (polarization degree was zero).

It is worth mentioning that variation of the polarization degree,
which we realize in our experiment, is impossible by means of only
linear lossless optical methods~\cite{pd}. However, in the present
experiment, the transformation performed over the signal photon is
essentially nonlinear since it is triggered by the detection of
the idler photon entangled to this photon.

On the other hand, it must be noticed that in this way the final
polarization state of the second photon depends on the quantum
efficiency of the first detector, since the Pockels cell is
activated only when the first photon is effectively detected. This
property, representing a limit for the operation of the
measurement-conditional unitary gate, gives nevertheless the
opportunity for realising an innovative scheme for absolute
calibration of single-photon detectors, which will be discussed
below.

\section{Experimental realization of the measurement-conditional
unitary gate}

In  our set-up (see Fig. 1,2) we have generated biphoton states by
pumping with an argon laser at 351 nm a BBO crystal (5x5x5 mm) cut
for producing type II Parametric Down Conversion. Residual pump
beam after passing through the non-linear crystal was absorbed by
a beam dumper.
 Then we have selected the directions where
the vertically polarized and horizontally polarized emitted
circles, corresponding to both photons having a 702 nm wavelength,
intersect. A quartz crystal follows the BBO crystal in order to
compensate the walk-off of ordinary and extraordinary rays due to
birefringence in BBO. In this way indistinguishability between the
two polarizations is restored generating the Bell state of Eq.
\ref{Psi}.

After crossing a polarizing beam splitter (PBS)  the first
(signal) photon was detected by means of a Perkin-Elmer single
photon detector preceded by a pinhole and a red filter. The second
photon (idler) was delayed by 200 m of optical fiber. Input and
output fiber coupling were realized with 20x objectives with a 0.4
numerical aperture. At the output of the fiber this photon was
then addressed to a KDP Pockels cell (supplied at 5.2 kV) followed
by a Glan-Thompson polarizer, an interference filter at 702 nm (4
nm FWHM) and a second Perkin-Elmer Silicon avalanche single photon
detector.

When the Pockels cell is not active the reduced polarization
density matrix of the second photon corresponds to a completely
unpolarized case. In fact, the observed intensity of the signal
measured when varying the setting of the Glan-Thompson polarizer
is flat (see Fig. 3).

On the other hand, when the Pockels cell is active the system
realizes a rotation of  the polarization of the second photon
conditioned to a polarization measurement on the conjugated arm.
If we choose to perform a rotation by $90^o$ in the same basis of
the polarization measurement of the first photon, the result is a
purification of the polarization state of the second one.

Our experimental results at coincidence level are shown in Fig. 4
in the case where we choose the $45^o -135^o$ basis both for
polarization measurement and polarization rotation (further
results will be presented in a following paragraph when discussing
detector calibration). In this basis the state is
\begin{equation}
\vert \phi ^- \rangle = {\frac{ \vert 45 \rangle \vert 45 \rangle
- \vert 135 \rangle \vert 135 \rangle }{\sqrt {2}}} \label{phi}
\end{equation}
As expected for an entangled state the maximum of coincidences is
shifted of $90^o$ when the Pockels cell is active: when the first
detector is preceded by a polarizer at $- 45^o$ the maximum of
coincidences is when the polarizer before the second detector is
set at $135^o$ with the Pockels cell off and at $45^o$ with the
Pockels cell on. The visibility is 87.2 \% when the Pockels cell
is off and 86.3 \% when it is on.

\section{ Description of the method for calibrating single photon detectors.}

This scheme for realizing a measurement conditional unitary gate
can be applied to the calibration of single photon-detectors. The
method is based on the fact that in the real situation the
efficiency $\eta_1$ of the first detector is smaller than unity
and thus only a fraction of incident photons will be observed and
produce an effect on the Pockels cell. Therefore the final density
matrix for the second photon does not correspond to the pure
state: \be \rho =  \vert V \rangle \langle V \vert \ee but to a
mixed one:
 \be
\rho =1/2 [ (1+\eta_1) \vert V \rangle \langle V \vert + (1-
\eta_1) \vert H \rangle \langle H \vert ] \ee which depends on the
quantum efficiency $\eta_1$ of the first detector, allowing a
calibration of it. If we insert a polarizer in front of the second
detector (Fig.1) and vary its angle $\theta$ (with respect to the
horizontal axis), the count rate $N_2$ of the second detector will
be given by: \be N_2=N_0 \cdot \alpha \eta_2 [ 1 - \eta_1 cos(2
\theta) ] \ee where $\alpha$ represents the idler optical path
loss (fiber and Pockels cell transmittances), $\eta_2$ is the
quantum efficiency of the second detector and $N_0$ the rate of
emission of entangled pairs.

The visibility $V$ of the signal counting rate $N_2$ obtained by
rotating the polarizer (G) preceding the second detector does not
depend on the quantum efficiency $\eta_2$ while it is directly
determined by the value of $\eta_1$,

\be V = { N^V_2 - N^H_2 \over N^V_2 + N^H_2 } = \eta_1 \ee
representing therefore a measurement of this ($N^H_2$ and $N^V_2$
are the counts on the second detector when the polarizer is set at
$0^o$ and $90^o$ respectively). No coincidence measurement is, in
principle, necessary.

Incidentally, this effect can be also described in terms of the
Stokes parameters, which are initially $S_0=N$, $S_1=S_2=S_3=0$,
where $N$ is the photon number, but become, as a result of
transformation, $S_0=N$, $S_1= \eta_1 N$, $S_2=S_3=0$. From the
Stokes parameters, one can find the polarization degree, which is
standardly defined as \be
P=\sqrt{\frac{S_1^2+S_2^2+S_3^2}{S_0^2}}.\ee We see that without
the transformation, $P=0$ and in the presence of the
transformation, $P=\eta_1$.

The same effect can also be described in terms of the von Neumann
entropy $S$, whose value characterizes the purity of a state,
going from zero for a pure state to unity for a completely mixed
one. In our case, if $\eta_1=1$ the final polarization reduced
density matrix $\rho _2$ of the second photon, corresponding to a
pure state, gives $S(\rho_2)=0$. On the other hand if $\eta_1=0$
the final polarization reduced density matrix of the second photon
is completely mixed with $S(\rho _2)=1$. Intermediate cases lie
between these two values (see Fig. 5).

\section{  Experimental results for the calibration scheme}

The set-up of the calibration scheme is substantially identical to
the one presented in section 3, only the idler photon is in this
case delayed by 50 m of single mode (4 $\mu m$ core) polarization
mantaining fiber. The detection apparatus driving the Pockels cell
(including filters and iris) is the device under calibration.
Nevertheless, in this case one does not need to have an entangled
state and the quartz crystal compensator is not necessary.
Therefore, for no compensator  introduced the produced state is a
mixed one:
\begin{equation}
\rho = {\frac{ \vert HV\rangle \langle HV \vert + \vert VH \rangle
\langle VH \vert }{2}}. \label{rho}
\end{equation}

Measurements are performed in the $0^o-90^o$ basis.

As before, when the Pockels cell in not activated the polarization
degree of the second photon is zero. When the Pockels cell is
active the system realizes a rotation of $90^o$ on the
polarization of the second photon conditioned to a measurement of
a vertically polarized photon on the conjugated arm. The results
of our measurement under  this condition are shown in Fig. 6 and
7. When the cell is active the measured signal has a $(1+ \eta_1
cos (2 \theta))$ behaviour as a function of the polarizer setting
$\theta$. When the background, estimated by rotating of $90^o$ the
pump laser polarization, is subtracted the data show a $(36.8 \pm
2.4) \%$ visibility.

This result represents a rough measurement of the quantum
efficiency that must be corrected for dead time of the system and
the efficiency of the Pockels cell.

The Pockels cell driver, when triggered by detector under
calibration, generates a high-voltage pulse with fast rising edge
(5 ns), a 100 ns flat-top and a  long fall tail of about 3.5 $\mu
s$ duration (Fig.8). When the mean trigger rate exceeds $10^4$
counts per second the Pockels cell driver is disabled for 1 s: in
order to make negligible this effect the counting rate on the
detector under calibration must be kept lower than the rate
threshold of $10$ kHz.

For the sake of completeness in Fig.9 we show the single counts on
the second detector for horizontal and vertical polarizations in
function of the delay and in Fig. 10 the same for coincidences. It
is clearly seen as the polarization of second photon is dominantly
vertical when the Pockels cell is on, whilst the two polarization
are equivalent, at single-count level, when the delay is too
large, so that the photon is received before the Pockels cell has
been activated. At coincidence level, when the Pockels cell is
activated, coincidences are observed for a vertical polarization.
When the delay is increased, the situation is reversed and
coincidence between vertically (first detector) and horizontally
polarized (second detector) photons is observed.

The effect of a finite efficiency of the Pockels cell apparatus
can be precisely estimated by measuring the visibility at
coincidence level (Fig.4). In fact in coincidences the effect of
the quantum efficiency of the detector under calibration is
irrelevant and the reduction of visibility is completely due to
the non-ideal Pockels cell apparatus. From the visibility at
coincidence level the efficiency of the Pockels cell apparatus can
be estimated to be $0.832 \pm 0.0023$ (see Fig.3). When this
further correction is introduced we have \be \eta_1 = { N_V - N_H
\over N_V + N_H } \cdot { N^c_V + N^c_H \over N^c_V - N^c_H } \ee
where $N_H$ and $N_V$ are the counts on the second detector when
the polarizer is set at $0^o$ and $90^o$ respectively and $N^c_H$
and $N^c_V$ the corresponding coincidence counts.
 From our data the measured quantum efficiency of the detection apparatus is
then $ 0.441 \pm 0.045$.

The uncertainties propagation formula can be written as
 \be
u^2(\eta_1) = c_1^2 u^2(N_H) + c_2^2 u^2(N_V) + c_3^2 u^2(N^c_H) +
c_4^2 u^2(N^c_V) \ee

 The $u$ are the uncertainties and the
sensitivity coefficients $c_i$ are evaluated by standard
propagation of uncertainties method.

A summary of the uncertainty budget is given in Table 1.

\begin{table}\begin{center}
\begin{tabular}{|l|r|l|r|l|r|}    %6 colonne left, right, left
\hline
\begin{scriptsize}
  Quantity \end{scriptsize}& \begin{scriptsize} Value \end{scriptsize} &  \begin{scriptsize}
  Standard Deviation \end{scriptsize}& \begin{scriptsize} Type of Distribution \end{scriptsize}&
\begin{scriptsize} Sensitivity Coefficient \end{scriptsize}& \begin{scriptsize} Uncertainty contribution
\end{scriptsize}
\\  \hline
$N_H$    &76.6& 4.2& Gaussian  & -0.006763 &   0.02840
\\  \hline
 $N_V$  &  165.9 & 5.7 & Gaussian & 0.003123 & 0.01780
\\  \hline
 $N^c_H$    & 4.4 & 1.6  & Gaussian & 0.01827 & 0.02923
\\  \hline
$N^c_V$    & 48.7 & 2.6 & Gaussian & -0.00165 & 0.00429

\end{tabular}

\caption{Uncertainty budget for single photon detector calibration
with the proposed scheme. Counts are per second.} \label{T1}
\end{center}
\end{table}

Finally, in order to compare with other calibration techniques, it
is worth introducing  a further correction due to losses in
polarizer cube. When this correction is made ($\epsilon = 0.9842$)
the final result is $\eta_1 = 0.448 \pm 0.045$ (if the correction
for a small drift of the pump power shift is made by taking into
account the change in the signal counts, the result becomes $0.454
\pm 0.032$ ). Incidentally, even a better precision can be
obtained by using a least square fit method applied to the data
presented in Fig.6, which leads to $\eta_1 = 0.486 \pm 0.011$
\cite{nosCPEM}.

For the sake of clarity, it must be emphasized again that the
reported quantum efficiency is not the "naked" detector one, but
the one corresponding to the detection apparatus including spatial
and spectral filtering. In many experimental situations this is
the datum necessary for understanding the performances of the
set-up. If one wants to measure the "naked" detector quantum
efficiency it would be necessary to introduce a corrective factor
keeping into account losses in the other elements in the detection
apparatus and in the non-linear crystal (corrections are of course
needed also for the other calibration method based on PDC). This
evaluation is beyond the purposes of this proof-of-principle
experiment.

Incidentally, for the sake of completeness, we would like to
report that similar results ($\eta = 0.417 \pm 0.024$), albeit
less accurate (even if a smaller statistical uncertainty) due to a
low efficiency of the Pockels cell, were also obtained by using a
Lithium Iodate Pockels cell.

  \section{ Comparison with the traditional calibration
scheme with biphotons}

In order to check the result obtained with the new scheme, we have
compared it with the traditional scheme of single photon detector
calibration by using biphotons.

 In more detail, this procedure \cite{bp1,alan}
consists of placing a couple of photon counting detectors
down-stream to the non-linear crystal, along the direction of
propagation of correlated photon pairs for a selected couple of
frequencies: the detection of an event on one detector guarantees
with certainty, thanks to the PDC biphotons properties, the
presence of a photon on the conjugated direction with a determined
wavelength. If N is the total number of photon pairs emitted from
the crystal in a given time interval $T_{gate}$ and $N_{signal}$,
$N_{idler}$ and $N_{coincidence}$ are the mean number of events
recorded, in the same time interval $T_{gate}$, by signal
detector, idler detector and in coincidence, respectively, we have
the following obvious relationships: \be N_{signal} = \eta_{
signal} N = \eta_{ idler} · N \ee where $\eta_{signal}$ and $
\eta_{idler}$ are the detection efficiencies on signal and idler
arms. The number of events in coincidence is \be N_{coincidence} =
\eta_{ signal} \eta_{idler}·N \ee due to the statistical
independence of the two detectors. Then the detection efficiency
$\eta_{signal}$ follows: \be \eta_{ signal} = N_{coincidence} /
N_{idler} \ee

This simple relation, slightly modified by taking into account
background subtraction and corrections for acquisition apparatuses
\cite{bp2}, is the base for the scheme for absolute calibration of
single photon detectors by means of PDC light, which reaches now
measurement uncertainty competitive  with traditional methods
\cite{alan,bp2}.

When we have applied this scheme to the same configuration as
described in the previous paragraph, our result has been $\eta =
0.4812 \pm 0.0015$, which includes the corrections \cite{bp2} for
detector dead time ($\tau=40 $ns) $\gamma = 1 - N_{signal} \tau$
and for the delay between start and stop signal in Time to
Amplitude Converter ($T= 9.3 $ns) $\alpha = 1 - N_{signal} T$. The
data from which this result is derived are shown in Fig.6 in
function of the counts on the detector under calibration (varied
by varying the power of pump laser) both with and without
corrections for detector dead time and for stop delay time.

For the sake of exemplification the uncertainty budget, for the
lower laser intensity, is reported in table 2.

This result is perfectly compatible with the one obtained with the
new scheme, $\eta_1 = 0.448 \pm 0.045$ (or $\eta_1 = 0.486  \pm
0.011$ with least squares fit method) . At the moment the
uncertainty is larger with the proposed scheme, but a substantial
reduction of this can be expected by a further careful
metrological analysis and determination of all the corrections and
an optimization of the Pockels cell apparatus.

Thus, these first results show that the proposed method could
allow an accuracy comparable with the existing ones and, in
particular, could be competitive with the one based on measurement
of coincidences. It is therefore worth of further accurate
metrological studies.

\begin{table}\begin{center}
\begin{tabular}{|l|r|l|r|l|r|}    %6 colonne left, right, left
\hline
\begin{scriptsize}
  Quantity \end{scriptsize}& \begin{scriptsize} Value \end{scriptsize} &  \begin{scriptsize}
  Standard Deviation \end{scriptsize}& \begin{scriptsize} Type of Distribution \end{scriptsize}&
\begin{scriptsize} Sensitivity Coefficient \end{scriptsize}& \begin{scriptsize} Uncertainty contribution
\end{scriptsize}
\\  \hline
$N_{i}$    & 1832.8& 9.0& Gaussian  & -0.00026 & 0.00234
\\  \hline
 $N_{c}$  &  874.4 & 5.2 & Gaussian & 0.000546&
 0.00284
\\  \hline
 $N_{s}$    & 131777 & 185  & Gaussian & $5.88 \cdot 10^{-10}$ &
 $1.1 \cdot 10^{-7}$
\\  \hline
T    & 9.3 ns & 0.5 ns & Rectangular & 1572 & $7.9 \cdot 10^{-7}$

\end{tabular}

\caption{Uncertainty budget for single-photon detector calibration
with traditional PDC scheme (point at lowest pump power). Counts
are per second.} \label{T1}
\end{center}
\end{table}

  \section{ Conclusions}

In conclusion we have described the realization of a
measurement-conditional unitary gate on biphoton states based on
the action of a Pockels cell on a member of a polarization
entangled pair of photons conditional to a polarization
measurement on the other member. This scheme can find various
application to Quantum Information processing \cite{NC} and
studies of Foundations of Quantum Mechanics \cite{au}.

 This unitary gate has then been
applied for giving a proof of principle of a new method for
absolute calibration of detectors based on rotation of
polarization of a member on a PDC biphoton state conditioned to
detection of the other member after polarization selection, since
 the polarization degree of the second state is given by the
quantum efficiency of the detector driving the Pockels cell.

These first results show that the proposed method could allow an
accuracy comparable with the existing ones and, in particular,
could be competitive with the one based on measurement of
coincidences. Of course, a further deep investigation of all the
details of this scheme and the use of a system realized on purpose
will be necessary for really reaching accuracy levels needed for
metrological applications.

It must be noticed that in principle this method could be used
even if the second detector is an analog one. Furthermore, we are
studying the possibility of extending this scheme for calibrating
analog detectors as well.

\vskip 0.5cm

{\bf Acknowledgments}

We acknowledge support of INTAS, grant \#01-2122.

One of us (L.Krivitsky) acknowledges the support of INTAS YS
fellowship grant (03-55-1971).

 Turin group acknowledges the support of MIUR (FIRB
RBAU01L5AZ-002; Cofinanziamento 2001) and Regione Piemonte.

 Moscow group acknowledges support of the Russian Foundation for Basic
Research, grant\#02-02-16664, and the Russian program of
scientific school support (\#166.2003.02).


\newpage
{\bf Figure Captions} \vskip 2cm
 Fig.1 Our set-up. A cw argon
laser generating at 351 nm pumps a type-II BBO crystal cut for
frequency-degenerate non-collinear polarization-entangled phase
matching. A quartz crystal compensate birefringence. One of the
correlated photons, after a spatial selection by means of a
pinhole A, a spectral selection by means of a red-glass cutoff
filter RG, and a polarization selection by means of a polarizing
cube PBS, is addressed to the photon counter D1, which drives,
through a fast high-voltage switch S, a Pockels cell PC placed in
the optical path of the other photon. The delay between a
photocount of D1 and the corresponding high-voltage pulse on the
Pockels cell can be varied electronically. The second photon of
the entangled pair is retarded, before the Pockels cell, by means
of  fiber F. This realizes the conditioned unitary operation. The
second photon is registered by photon counter D2 preceded by a
Glan prism G and an interference filter IF. The output signals
from the detectors are routed  to a two channel counter C, in
order to have the number of events on single channel, and to a
Time to Amplitude Converter circuit, followed by a single channel
analyzer, for selecting and counting coincidence events.

Fig. 2  Picture of our optical bench. One can recognize the two
detectors preceded by
   filters,iris and polarizers,
   the Pockels cell and the fiber.

Fig. 3 Counting rate of the second detector as a function of the
angle of the polarizer preceding it (without background
subtraction) in absence of the Pockels cell.

Fig. 4 Coincidences as a function of the angle of the polarizer
preceding the second detector for the Bell state $\vert \phi ^-
\rangle = {\frac{ \vert 45 \rangle \vert 45 \rangle - \vert 135
\rangle \vert 135 \rangle }{\sqrt {2}}} \label{phi}$. When the
Pockels cell (KDP) is not activated the maximum is at $135^o$
(squares). When a $90^o$ rotation of polarization is realized by
the Pockels cell conditioned to the measurement of a $45 ^o$
polarized photon in the conjugated arm, the maximum is shifted, as
expected, to $45^o$ (triangles).

Fig. 5 Von Neumann entropy of the trigger photon polarization
density matrix
   as a function of the quantum efficiency $\eta_1$ of the first detector. When $\eta_1=1$ the final
polarization reduced density matrix $\rho _2$ of the second photon
corresponds to a pure state, giving $S(\rho_2)=0$. On the other
hand if $\eta_1=0$ the final polarization reduced density matrix
of the second photon is completely mixed with $S(\rho _2)=1$.
\label{S}

Fig. 6 Counts on the second detector as a function of the angle of
the polarizer preceding it (without background subtraction). When
the Pockels cell (KDP) is not activated no dependence on the
polarizer angle appears (squares). When a $90^o$ rotation of
polarization is realized by the Pockels cell conditioned to the
measurement of a vertically polarized photon on the conjugated
branch, the data (triangles) show a clear dependence on the
polarizer setting (corresponding to a mainly vertically polarized
state).

Fig. 7  Coincidences as a function of the angle of the polarizer
preceding the second detector for the state described by the
density matrix $\rho =1/2 ( \eta_1 \vert V \rangle \langle V \vert
+ (1- \eta_1) \vert H \rangle \langle H \vert)$ . When the Pockels
cell (KDP) is not activated a maximum is at $0^o$ (squares). When
a $90^o$ rotation of polarization is realized by the Pockels cell
conditioned to the measurement of a vertically polarized photon on
the conjugated branch, the maximum is shifted, as expected, by
$90^o$ (triangles).

Fig. 8  The pulse generated by the Pockels cell drive and the
light measured after the Pockels cell between crossed polarizers
when activated by the former pulse and reached by a He-Ne laser
beam.

Fig. 9 Dependence of the D2 counting rate at $\theta=0^o$
(Horizontally polarized photons, triangles) and $\theta=90^o$
(Vertically polarized photons, squares) on the delay $T$
introduced electronically between the trigger pulses from D1
detector and the corresponding high-voltage pulses driving the
Pockel's cell. The effect of counting rate decreasing for
horizontally polarized photons is observed for  $T < 100 \, ns$
and not observed for $T>100 \, ns$.

Fig. 10 Dependence of coincidences at $\theta=0^o$ (Horizontally
polarized photons, triangles) and $\theta=90^o$ (Vertically
polarized photons, squares) for the polarization selection of the
second photon on the delay $T$ introduced electronically between
the trigger pulses from D1 detector and the corresponding
high-voltage pulses driving the Pockel's cell. The effect of
reversing of coincidences between H and V selection  is observed
at $T < 100 \, ns$.

Fig. 11  Coincidences  in function of the counts on the detector
under calibration (by varying the power of pump laser) both with
(circles) and without (squares) corrections for detector dead time
and for stop delay time \cite{bp2}. These data are used for
evaluating the quantum efficiency of the signal detector by means
of traditional scheme for biphoton calibration of detectors
\cite{bp1,alan}.

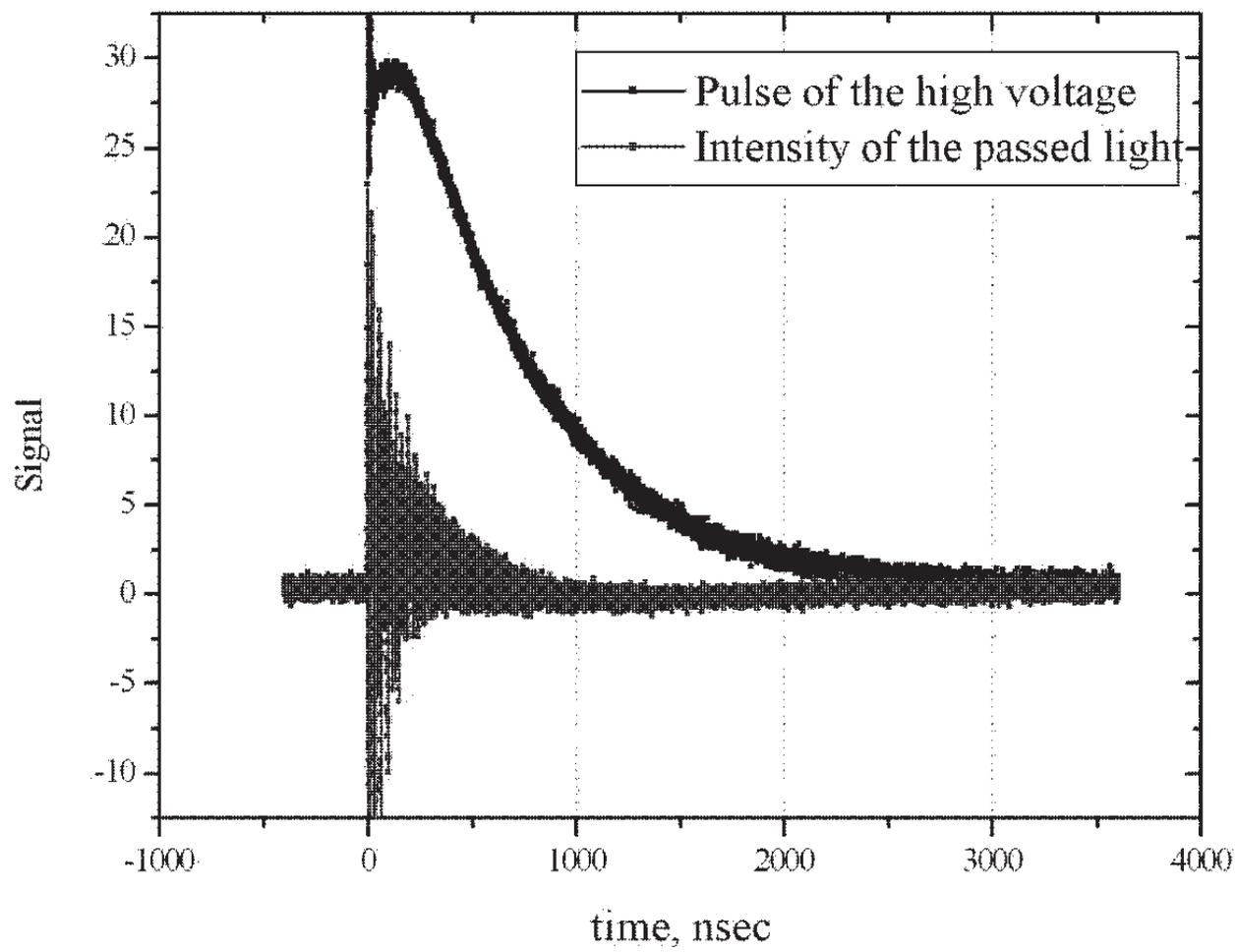
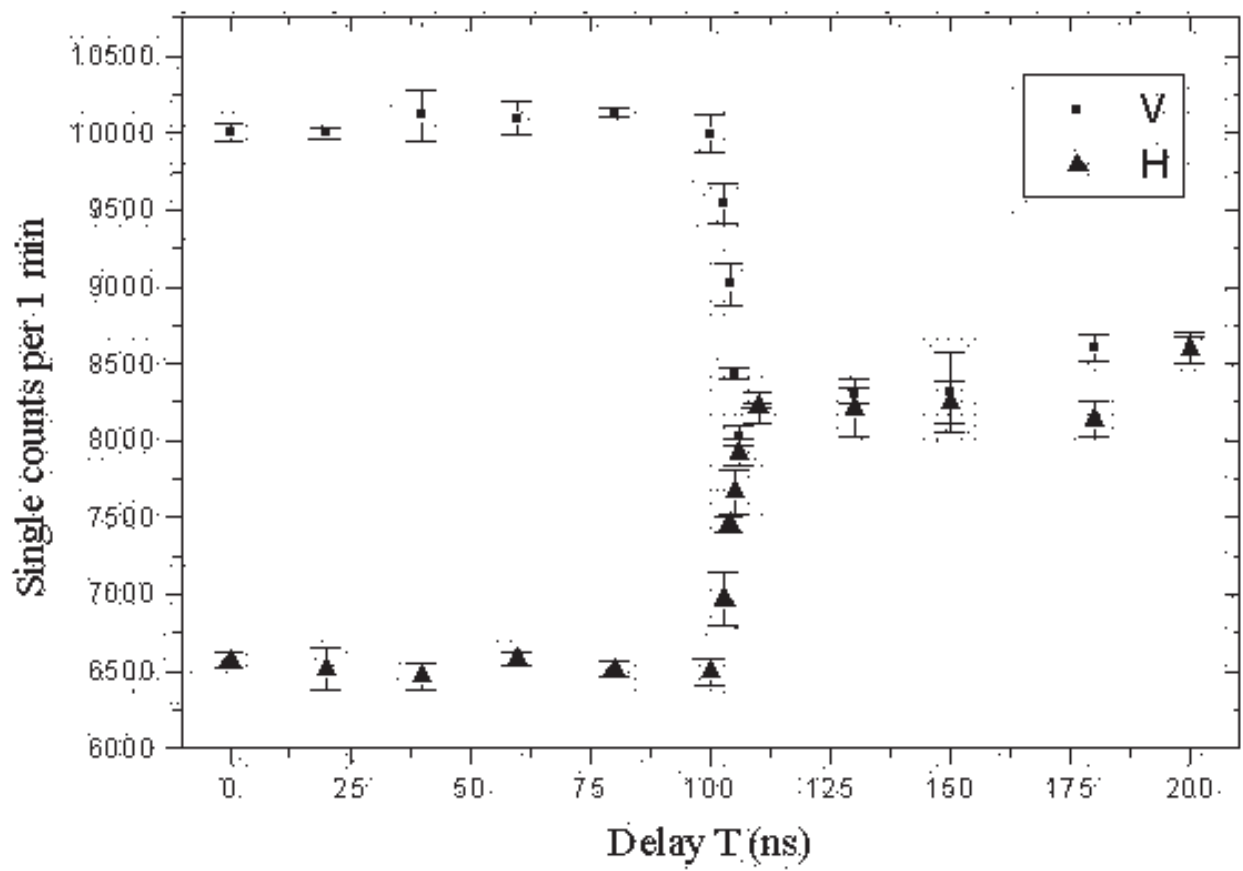
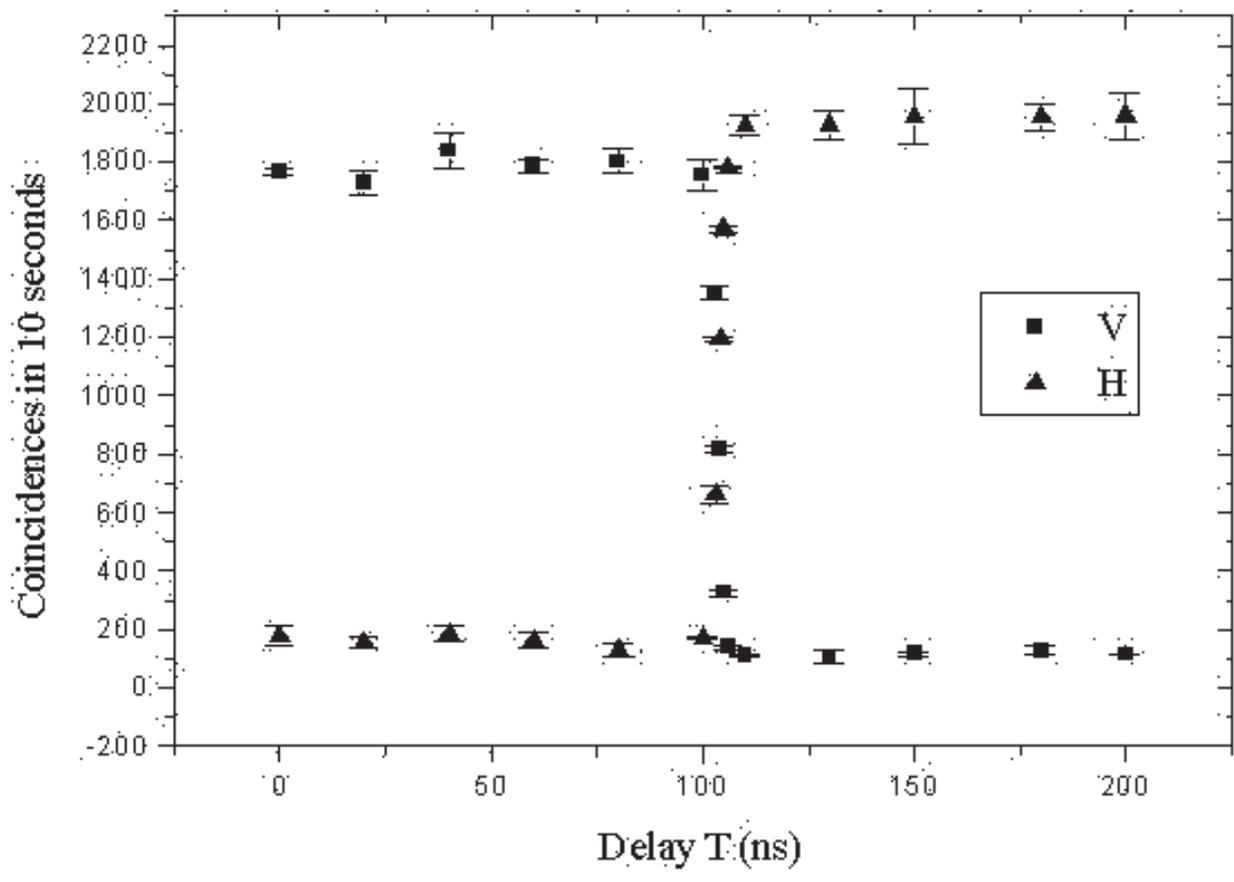
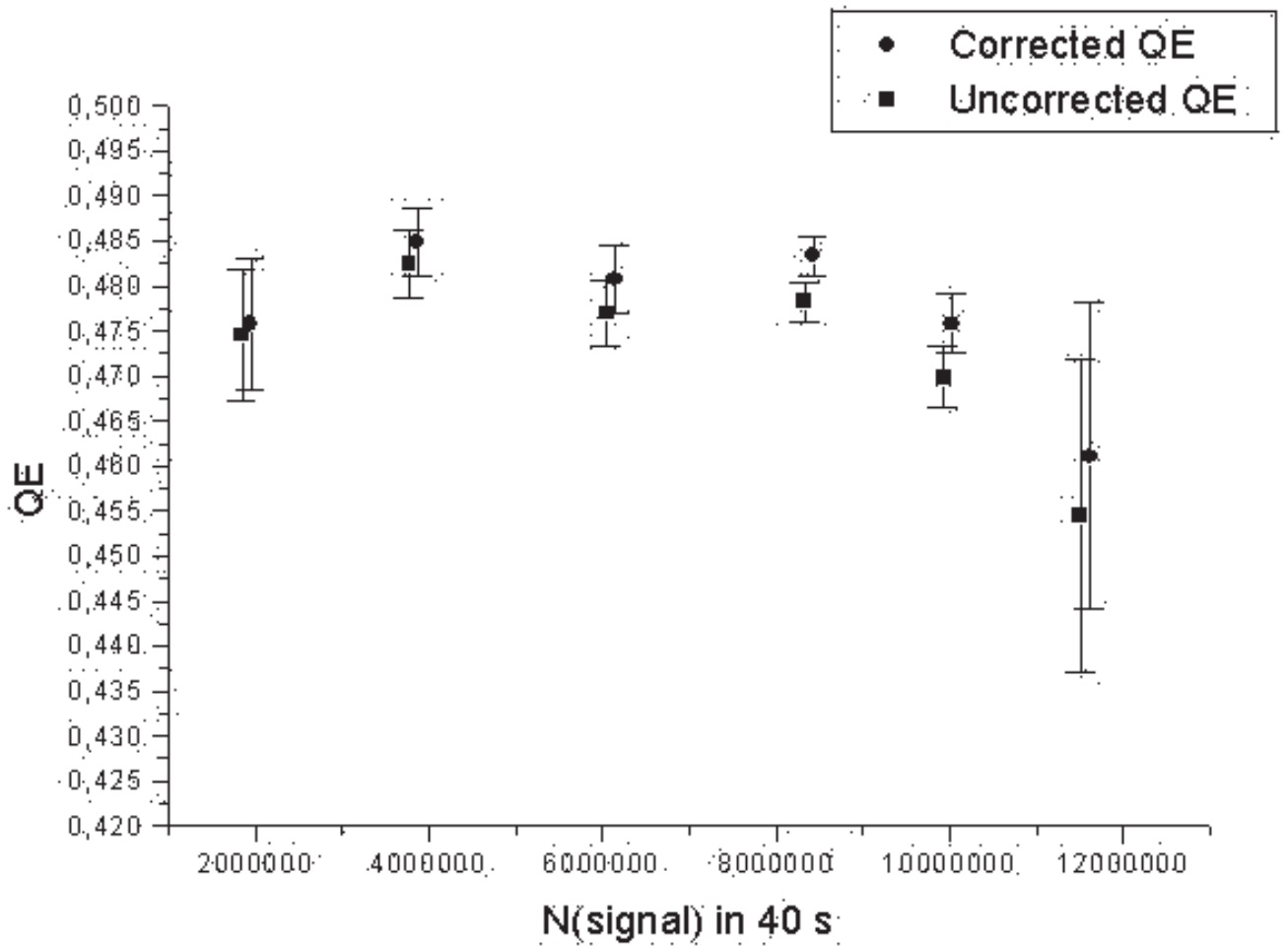
\end{document}